\documentclass[preprint2]{aastex}

\shorttitle{Massive Black Holes in Star Clusters}
\shortauthors{Baumgardt, Makino, Ebisuzaki}

\begin{document}

\title{Massive Black Holes in Star Clusters. II. Realistic Cluster Models}
\author{Holger Baumgardt\altaffilmark{1},
        Junichiro Makino\altaffilmark{2},
        Toshikazu Ebisuzaki\altaffilmark{1}}

\altaffiltext{1}{
        Astrophysical Computing Center, RIKEN, 2-1 Hirosawa, Wako-shi,
        Saitama 351-0198, Japan}

\altaffiltext{2}{
        Department of Astronomy, University of Tokyo, 7-3-1 Hongo,
        Bunkyo-ku,Tokyo 113-0033, Japan}

\begin{abstract}
We have followed the evolution of multi-mass star clusters containing massive central black holes through
collisional $N$-body simulations done on GRAPE6. Each cluster is composed of between
16,384 to 131,072 stars together with a black hole with an initial mass of $M_{BH}=1000 M_\odot$. 
We follow the evolution
of the clusters under the combined influence of two-body relaxation, stellar mass-loss and tidal disruption
of stars by the massive central black hole.

We find that the (3D) mass density profile follows a power-law distribution $\rho \sim r^{-\alpha}$ with
slope $\alpha=1.55$ inside the sphere of influence of the central black hole. This leads to a constant
density profile of bright stars in projection, which makes it highly unlikely that core collapse clusters 
contain intermediate-mass black holes. Instead globular clusters containing massive central black 
holes can be fitted with standard King profiles. Due to energy generation in the
cusp, star clusters with intermediate-mass black holes (IMBHs) expand. The cluster expansion is so
strong 
that clusters which start very concentrated can end up among the least dense clusters.
The amount of mass segregation in the core is also smaller
compared to post-collapse clusters without IMBHs. 
Most stellar mass black holes with masses $M_{BH}>5 M_\odot$ are lost from the clusters within
a few Gyrs through mutual encounters in the cusp around the IMBH. 

Black holes in star clusters disrupt mainly main-sequence stars and giants and no neutron stars. The disruption rates 
are too small to form an IMBH out of a $M_{BH} \approx 50$ $M_{BH}$ progenitor black hole even if all material
from disrupted stars is accreted onto the black hole, unless star clusters start with central densities significantly
higher than what is seen in present day globular clusters. 

We also discuss the possible detection mechanisms for intermediate-mass black holes. Our simulations show 
that kinematical studies can reveal 1000 $M_\odot$ IMBHs in the closest clusters.
IMBHs in globular clusters are only weak X-ray sources since the tidal disruption rate
of stars is low and the star closest to the IMBH is normally another black hole which
prevents other stars from undergoing stable mass transfer.
For globular clusters, dynamical evolution can push compact stars 
near the IMBH to distances small enough that they become detectable sources of gravitational radiation.
If 10\% of all globular clusters contain IMBHs, extragalactic globular clusters could be one of the 
major sources of gravitational wave events for {\it LISA}.
\end{abstract}


\newcommand{\msun}{M_{\odot}}

\section{Introduction}

This is the second paper in a series of $N$-body simulations which deal with the dynamical evolution of star 
clusters containing
massive central black holes. In \citet{Baumgardtetal2004a} (paper I), we followed the evolution 
of single mass clusters with black holes that contained a few percent of the total
system mass. We showed that the density distribution of stars inside the sphere of influence of the black
hole follows a $\rho \sim r^{-1.75}$ power law, in good agreement with results from previous
analytical estimates and indirect simulation methods \citep{BahcallWolf1976, CohnKulsrud1978,
MarchantShapiro1980}. We 
also derived the rate of tidal disruption of
stars. The present paper extends these results to
the more realistic but much less studied case of a star cluster with a spectrum of stellar masses 
and is aimed at 
globular clusters with intermediate-mass black holes of a few hundred to a few thousand
solar masses at their centres.

X-ray observations of starburst and interacting galaxies have revealed a class of ultra-luminous X-ray
sources (ULX), whose luminosities exceed the Eddington luminosities of stellar mass black holes
by orders of magnitude \citep{Makishimaetal2000}, making them good candidates for IMBHs. 
The irregular
galaxy M82, for example, hosts an ULX with luminosity
$L > 10^{40}$ ergs/sec near its centre \citep{Matsumotoetal2001, Kaaretetal2001} the position
of which coincides with that of the young ($T \approx 10$ Myrs) star 
cluster MGG-11. \citet{PortegiesZwartetal2004} have performed $N$-body simulations of several star clusters in M82,
using the cluster parameters determined by \citet{McCradyetal2003}. They found that runaway merging 
of massive stars could have led to the formation of an IMBH with a few hundred to a few thousand
solar masses in MGG-11, thereby explaining the presence of the ultraluminous X-ray source.
General conditions when runaway merging of stars can lead to the formation of IMBHs were
discussed in \citet{PortegiesZwartMcMillan2002} and \citet{Rasioetal2004}. 

Apart from run-away collisions of main sequence stars, massive black holes could
also be build up through the merger of stellar mass black holes \citep{MouriTaniguchi2002} via
gravitational radiation in dense enough star clusters. In less concentrated clusters,
this process is also possible but may take up to a Hubble time \citep{MillerHamilton2002}.
Finally, intermediate-mass black holes could also form by the accretion 
onto a stellar-mass black hole of interstellar cluster gas 
due to radiation drag from bright stars, provided the stellar
mass function of the cluster is shallow enough (Kawakatu \& Umemura 2004).

Although there are several possible ways to form IMBHs, 
the observational evidence for intermediate-mass black holes in star clusters is as yet 
much less clear.
Due to their high central densities and since they are relatively close, galactic core-collapse clusters 
are obvious places to look for intermediate-mass black holes. Indeed, for almost 30 years, M15
has been thought to harbour an intermediate-mass black hole in its centre \citep{Newelletal1976}.
The
most recent analysis of this cluster was done by \citet{Gerssenetal2002, Gerssenetal2003}, who
used HST to obtain new spectra of stars in the central cluster region. They found that the 
velocity dispersion can best be
explained by an intermediate-mass black hole of mass $ (1.7 \pm 2.7) \cdot 10^3 M_\odot$. However, 
\citet{Baumgardtetal2003a} have shown that the observational data can also be explained by the core-collapse profile 
of a 'standard' star cluster without a massive central black hole. In this case the central rise of 
the mass-to-light ratio is created by an
unseen concentration of neutron stars and heavy mass white dwarfs. Such a model is also able to explain
the velocity dispersion derived from the proper motion of stars near the centre of M15
\citep{McNamaraetal2003, McNamaraetal2004}.
Similarly, a dense concentration of compact remnants 
might also be responsible for the high mass-to-light ratio of the central region of NGC 6752 seen in
pulsar timings \citep{Ferraroetal2003, Colpietal2003}.
Outside our own galaxy, \citet{Gebhardtetal2002} have 
reported evidence for a 20,000 $M_\odot$ black hole in the M31 globular cluster G1, but 
\citet{Baumgardtetal2003b} showed that dynamical simulations without black holes completely explain
the observed velocity dispersion and density profile of this cluster.

Despite the unclear observational situation, the presence of intermediate-mass black holes in star
clusters remains an interesting possibility. They would provide the missing link between the stellar
mass black holes which form as a result of stellar evolution and the supermassive black holes
in galactic centres \citep{Ebisuzakietal2001}. They would also be prime targets for the forthcoming
generation of both ground and space-based gravitational wave detectors due to their high masses and
the fact that they have the potential to merge with other black holes if residing in dense star clusters.

In this paper we study the dynamical behaviour of massive black holes in globular clusters. Section 2
describes the set-up of our runs and in section 3 we present our main results concerning the dynamical evolution
of the star clusters, the tidal disruption of stars and the possibilities of detecting the central
black hole. In section 4 we report our conclusions.

\section{Description of the runs}

We simulated the evolution of clusters containing between $N = 16,384$ and $131,072$ (128K) stars
using Aarseth's collisional $N$-body code NBODY4 \citep{Aarseth1999} on the GRAPE6 computers of Tokyo 
University \citep{Makinoetal2003}. Most clusters were treated as isolated and followed King $W_0=7.0$ 
profiles initially. Simulations were run for a Hubble time, which was assumed to be $T=12$ Gyrs.
Since black holes probably form early on in the evolution of a globular cluster by e.g.\ run-away 
merging of massive stars, we started our runs with a massive black 
hole at rest at the cluster centre. As in paper I, we modified the velocities of the cluster 
stars to prevent the cluster centre from collapsing after adding the IMBH to the cluster.

All clusters started with a central black hole of mass $M_{BH} = 1000 M_\odot$. If the 
$M_\odot - \sigma$ relation
found by \citet{Gebhardtetal2000} for galactic bulges holds for globular clusters as well,
this corresponds to the IMBH mass expected in a typical globular cluster. In addition, the best case
found so far for an IMBH in a star cluster, M82 X-1 in MGG-11, must have a mass between a few hundred to a 
few thousand solar masses based on its X-ray luminosity and the frequency of quasi-periodic
oscillations seen in the X-ray flux \citep{Matsumotoetal2001, StrohmayerMushotzky2003}. 

Stellar evolution was treated by the fitting formulae of \citet{Hurleyetal2000},
assuming a mean cluster metallicity of
$Z=0.001$ and a neutron star retention fraction of 15\%. The 15\% retention fraction was imposed
by immediately removing 85\% of all neutron stars at the time of their formation while leaving the
velocities of the remaining ones unchanged. The exact form of the 
IMF at the low-mass end will not critically influence the
results of our simulations as such stars are mere test particles in the gravitational field of the
higher mass stars. More important is the mass function at the high-mass end, especially the number and
mass distribution of black holes formed during the run. Unfortunately, the initial-to-final mass relation 
for high-mass stars is currently not precisely
known since it depends among other things on the assumed amount of stellar wind mass-loss in the final 
phases of stellar evolution and the details of the explosion mechanism \citep{FryerKalogera2001}.
In addition the metallicity of the progenitor star will effect the mass of the final black hole 
\citep{Hegeretal2003} for low-metallicity stars, so globular clusters with different metallicities
might have different black hole mass distributions. Since the fraction and mass distribution
of heavy mass black holes could have a strong influence on the outcome of our simulations, we performed
two series of simulations.

\begin{table*}
\caption[]{Details of the performed $N$-body runs.}
\begin{tabular}{rcrr@{.}lr@{.}lccccr}
\noalign{\smallskip}
 \multicolumn{1}{c}{$N$}& \multicolumn{1}{c}{$N_{Sim}$} & 
\multicolumn{1}{c}{$W_0$} & \multicolumn{2}{c}{$M_{up}$} & 
 \multicolumn{2}{c}{$M^1_{CL}$} &
\multicolumn{1}{c}{$r_{h\,i}$} & \multicolumn{1}{c}{$r^1_{h\,f}$} & 
  \multicolumn{1}{c}{$M_{BH\,i}$} & \multicolumn{1}{c}{$M^1_{BH\,f}$} & 
  \multicolumn{1}{c}{$N^1_{Tid}$}\\
   & & &  \multicolumn{2}{c}{[$M_\odot$]} & \multicolumn{2}{c}{[$M_\odot$]}  & [pc] & 
     [pc] &  [$M_\odot$] &  [$M_\odot$] & \\[+0.1cm]
  16384 & 4 & 7.0 & 30 & 0 &  9778 & 5 & 4.87 & 28.06 & 1000.0 & 1007.2 &  8\\
  32768 & 2 & 7.0 & 30 & 0 & 18809 & 5 & 4.87 & 21.97 & 1000.0 & 1023.7 & 19\\
  65536 & 1 & 7.0 & 30 & 0 & 39310 & 9 & 4.87 & 17.33 & 1000.0 & 1030.3 & 27\\
 131072 & 1 & 7.0 & 30 & 0 & 76936 & 8 & 4.87 & 13.98 & 1000.0 & 1045.5 & 37\\
\noalign{\smallskip}
  32768 & 2 & 7.0 & 100 & 0 & 20681 & 2 & 4.87 & 27.39 & 1000.0 & 1003.3 &  2\\
  65536 & 1 & 7.0 & 100 & 0 & 41024 & 1 & 3.86 & 18.97 & 1000.0 & 1001.9 &  5\\
 131072 & 1 & 7.0 & 100 & 0 & 83919 & 4 & 3.07 & 14.00 & 1000.0 & 1004.4 & 11\\
\noalign{\smallskip}
  16384 & 2 & 7.0 &  30 & 0 &  9739 & 6 & 0.79 & 32.06 & 1000.0 & 1039.9 & 24\\
  16384 & 2 &10.0 &  30 & 0 &  9868 & 9 & 6.23 & 29.40 & 1000.0 & 1015.8 & 12\\
\end{tabular}
\begin{flushleft}
Notes:  \\

1. For clusters with more than one simulation, parameters given are average values.
\end{flushleft}
\end{table*}

In our first series of simulations, the mass function of the cluster stars was given by a \citet{Kroupa2001} 
IMF with a lower mass limit of $0.1 M_\odot$ and an upper mass limit of $30 M_\odot$.
Since this upper mass limit is only slightly above
the mass where black holes instead of neutron stars form due to stellar evolution, these clusters 
contain only a small fraction of stellar mass black holes, all of them with masses below $3 M_\odot$.
In the second series of simulations we used an upper mass limit of 100 $M_\odot$ and transformed the
stars directly into black holes without further mass-loss. In this series, the most
heavy mass black holes formed have about 45 $M_\odot$. In both types of simulations, all black holes 
were retained in the clusters. These two cases are the most extreme models, and most likely real
globular clusters fall in between.

We did not include a primordial binary population. The presence of a
primordial binary population might help the formation of IMBH through
enhancing the stellar collisions (e.g. \citet{Fregeauetal2003}). Its effect
on the structure of a cluster with central IMBH would be to decrease
the central density through hardening. Thus, our present result
without primordial binaries gives the upper limit of the central
density. 

Stars were assumed tidally disrupted if their distance to the central black hole was smaller than 
the critical distance given by eq. 3.2 in \citet{Kochanek1992}:
\begin{equation}
r_t = 1.3 \left( \frac{M_{BH}}{2 \, M_*} \right)^{1/3} R_* \;\; ,
\label{koch}
\end{equation}
where $M_{BH}$ and $M_*$ are the mass of the black hole and the star and $R_*$ is the
stellar radius. The stellar radii 
were taken from the formulae of \citet{Hurleyetal2000}. 
The mass of tidally disrupted stars was added to the mass of the central black hole.
So far we have not incorporated the 
effects of gravitational radiation in our runs since the central densities reached in our simulations
are not large enough that gravitational radiation becomes important. 
We also did not include stellar collisions, which become dynamically important when the 
velocity dispersion around the black hole becomes equal to the escape velocity from the
stellar surface, as stars cannot undergo large angle encounters any more \citep{FrankRees1976}. Even 
for main sequence stars,
this corresponds to distances of $r_{coll}=10^{-5}$ pc from the black hole, which is far inside the distance
of the innermost stars from the IMBH.

Table 1 gives an overview of the simulations performed. It shows the number of cluster
stars $N$, the number of simulations $N_{Sim}$ performed for a given model, the initial concentration $W_0$
of the King model, the upper mass limit of the IMF, 
initial cluster mass and half-mass radius. Also shown are the final half-mass radius, and the black hole masses
at the start and the end of the runs. The final column gives the (average) number of tidal disruptions.

\section{Results}

\subsection{Cluster expansion}

\begin{figure}[tb!]
\plotone{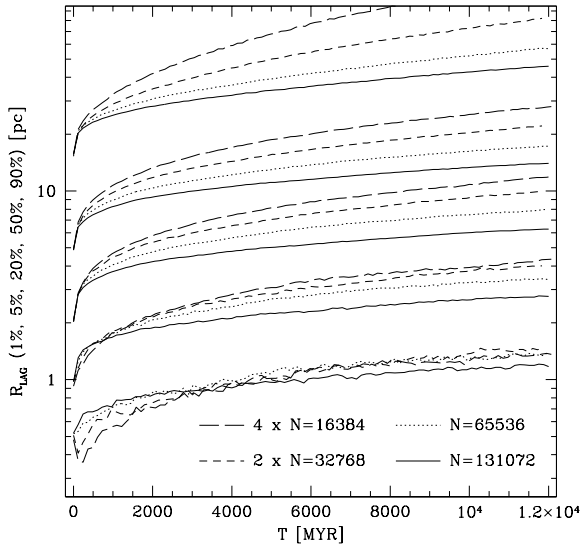}
\caption{Lagrangian radii as a function of time for the first four cluster simulations with particle 
 numbers between
 $16,384 \le N \le 131,072$. All clusters expand as a result of energy generation in the cusp around the
 black hole and initially also due to stellar evolutional mass-loss which decreases the binding
 energy of the cluster. Since two-body relaxation drives the expansion, high-$N$ models expand less
  than low-$N$ ones.}
\label{cl_exp}
\end{figure}

Fig.\ \ref{cl_exp} depicts the evolution of lagrangian radii for the first four clusters of Table 1. 
All clusters expand since mass-loss of individual stars due to stellar evolution decreases  
the potential energy of the clusters and two-body processes in the cusp around the black hole exchange 
energy between the stars. As a result, the innermost stars are pushed to more negative energies until
they are tidally disrupted by the IMBH while the rest of the cluster stars gain energy and the cluster
expands. This behaviour is similar to the one seen in the single mass runs of paper I (Fig.\ 5).

For all clusters the expansion is strongest in the initial phase since the cluster radii and
therefore the two-body relaxation time is smallest in the beginning. In addition, the mass-loss of stars due
stellar evolution is strongest within the first Gyr. The expansion is smaller for high-$N$ clusters
due to their longer relaxation times.
Table 1 shows that for a given mass of the central 
black hole, the final half-mass radius depends only on the number of cluster stars and is nearly
independent of the cluster composition and initial half-mass radius. For the N=16K clusters for example, the
final half-mass radius changes by less than 10\% if the initial half-mass radius is reduced by a
factor of 6. The final radius also does not depend much on the initial IMF or the initial concentration 
of the King model. 

The two-body relaxation time of a
cluster with mass $M_C$ and radius $r_h$ is given by \citep{Spitzer1987}:
\begin{equation}
  T_{RH} \sim \frac{\sqrt{M_C} \; r_h^{3/2}}{<\!m\!> \; \sqrt{G} \; \ln (\gamma N)} \;\; ,
\label{trh}
\end{equation}
for a cluster with stars of average mass $m$. If the cluster expansion is caused by
two-body relaxation, half-mass radii of clusters with masses $M_C$ should satisfy a relation 
$r_h \sim M_C^{-1/3}$ after a long enough time has passed since the half-mass relaxation 
time is the same for all clusters in this case so they expand with the same rate. 
The final half-mass radii of the clusters in Table 1 are in good agreement with such a scaling law.
\begin{figure}[tb!]
\plotone{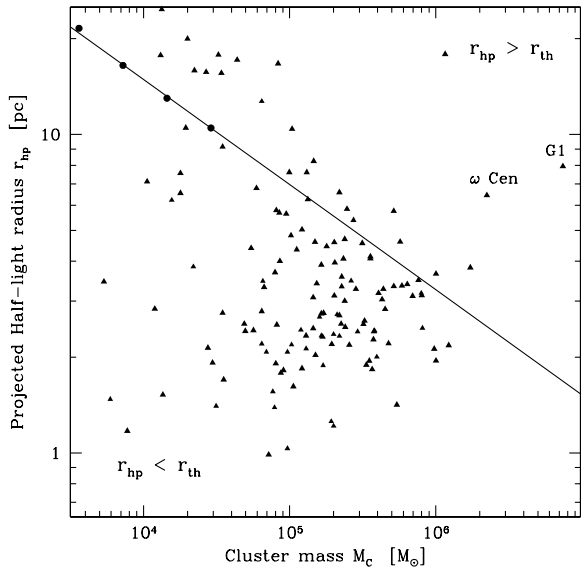}
\caption{Projected half-light radii of globular clusters $r_{hp}$ against their mass (triangles).
Projected half-light radii of the clusters in the $N$-body simulation are marked
by filled circles. They follow a relation $r_{hp} \sim M_C^{-1/3}$ (solid line), in agreement with the
idea that two-body relaxation drives the cluster expansion. 
The expansion due to an IMBH is strong enough that clusters with IMBHs can end up among the 
least concentrated clusters.}
\label{radcmp}
\end{figure}

Fig.\ \ref{radcmp} compares the projected half-light radii of the clusters in Fig.\ \ref{cl_exp}
with the projected half-light radii of galactic globular clusters. The results of the $N$-body
simulations are well fitted by an $r_{hp} \sim M_C^{-1/3}$ law (solid line), so projected half-mass
and half-light radii follow the same relation as the 3D ones. Projected half-light
radii of galactic globular clusters are taken from \citet{Harris1996}. Masses for globular clusters
were calculated from their absolute $V$ magnitudes, assuming a mass-to-light ratio of $m/L_V=2.0$.
Most globular clusters have projected half-light
radii which are smaller than the ones predicted by an extrapolation of our runs. 

\citet{Rasioetal2004} and \citet{PortegiesZwartetal2004} have shown that core-collapse times of less than a Myr
are necessary to form an IMBH in the centre of a star cluster by run-away collision of main sequence stars.
For high-$N$ clusters this corresponds to central densities of $\rho_c=10^6 M_\odot/$pc$^3$ or higher.
Most other processes which have been proposed as formation
mechanisms for IMBHs also require high density environments.
Such densities are among the highest found in globular clusters (see Table 2 of Pryor \& Meylan 1993).
Figs.\ \ref{cl_exp} and \ref{radcmp} show that even if clusters with massive black holes
start with very high densities, the subsequent cluster expansion increases their radii by
such an amount that they can end up among the least dense clusters. Thus, the low density of 
present-day globular clusters do not rule out the formation of IMBH, since they might have 
been much more compact when they were born.

The fact that most clusters have half-light radii below our prediction does not
speak against the presence of IMBHs in these clusters, since, while our
clusters were isolated, galactic globular clusters are surrounded by a tidal field, which limits
their growth. Tidally limited clusters which lie below our predicted line might
therefore still contain IMBHs. 

The half-mass radius of G1, the most massive globular cluster in M31, is $r_h=8$ pc 
\citep{Baumgardtetal2003b}. If G1 started much more concentrated and its current size is due to an 
expansion similar to the one seen in our runs, the mass of the central IMBH must be larger than
1000 $M_\odot$ since the half-mass radius of G1 is much higher than predicted by our runs.
The same is true for $\omega$ Cen, the most massive galactic globular
cluster.

\subsection{Density profile}

In paper I it was shown that the density profile of a single-mass cluster follows a power-law profile
$\rho \sim r^{-\alpha}$ inside the influence radius of the black hole with slope $\alpha=1.75$,
in agreement with results from Fokker-Planck and Monte Carlo simulations. It was also shown that the 
sphere of influence 
of the black hole is limited by two conditions. For small-mass black holes, clusters have a near
constant density core outside the sphere of influence of the black hole and the central cusp extends 
only up to a radius $r_i$ at which the 
velocity dispersion in the core becomes comparable to the circular velocity of stars around
the black hole. The following relation was found to give a good estimate for $r_i$: 
\begin{equation}
 r_i = \frac{G M_{BH}}{2 \, <\!v_c^2\!>} \;\; ,
 \label{ri}
\end{equation}
where $v_c$ is the core velocity dispersion. For black holes which contain more than a few percent of the
cluster mass, a second condition for $r_i$ was found to be that the mass in stars inside $r_i$
must be smaller
than the mass of the central black hole, since otherwise the self-gravity of stars becomes
important and changes the density law. 

\begin{figure}[tb!]
\plotone{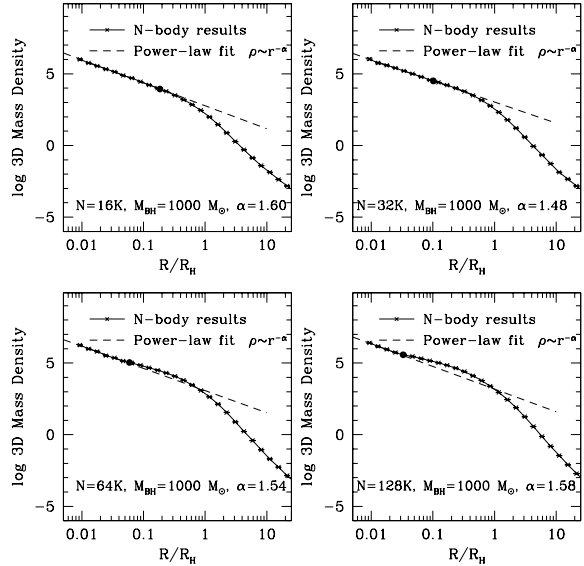}
\caption{3D mass density profile after T=12 Gyrs for 4 cluster simulations starting with particle 
numbers between
 $16,384 \le N \le 131,072$. Solid lines mark the $N$-body results, dashed lines
   a single power-law fit to the density profiles inside the radius of influence of the black hole 
    (shown by
     a solid circle). For all models we obtain slopes near $\alpha=1.55$ for the central stellar cusp.}
\label{3d_dens}
\end{figure}

Fig.\ \ref{3d_dens} depicts the final density profile of the first four clusters after 12 Gyrs. Shown is
the three-dimensional mass density of all stars. In order to calculate the density profile, we have 
superimposed between 5 (128K) to 20 (16K) snapshots centered at T=12 Gyrs, creating roughly the same 
statistical
uncertainty for all models. All snapshots were centered on the position of the IMBH.
We then fitted the combined density profile inside the influence radius of the black hole
with a power-law density profile.
It can be seen that we obtain a power-law profile inside $r_i$ with
slope around $\alpha=1.55$. There is no visible dependence on the
particle number. It will be shown in section 3.3,
where the mass segregation in the clusters is discussed, that the most massive stars in our
clusters still follow an $\alpha=1.75$ power-law, but that they are not numerous enough to
dominate the central region which is the reason for the flatter overall slope seen in our runs.

For $N=16$K, the black hole contains more than 10\% of the total cluster mass at $T = 12$ Gyrs and 
it dominates the density profile throughout the core.
For clusters with particle numbers more realistic for globular clusters, the central black 
holes contains an increasingly smaller fraction of the total cluster mass, so the velocity criterion
limits the influence of the black hole. In this case the central cusp contains only a fraction of the 
mass of the central black hole, for $N=128$K for example only about 10\%, i.e.\ $100 M_\odot$. Since
a considerable fraction of these stars would not be easily visibly to an observer because they
are compact remnants and therefore too faint, the direct observation of this cusp for globular
clusters with $M_{BH}<1000M_\odot$ IMBHs
is nearly impossible due to statistical uncertainties.
\begin{figure}[tb!]
\plotone{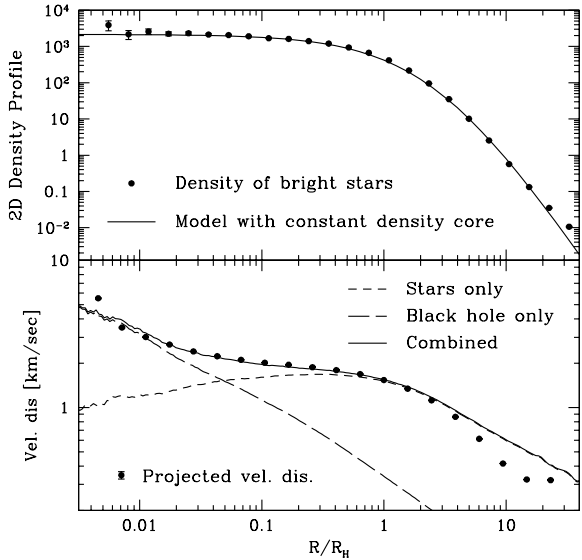}
\caption{Projected density profile of bright stars (top) and projected velocity dispersion of the cluster
 simulation starting with $N=131,072$ stars. The projected distribution of bright stars has a constant
 density core, similar to that seen in most globular clusters.
  Observations of the velocity dispersion could reveal the black hole if a sufficiently large number of
   stars at radii $r/r_h<0.01$ can be observed (bottom panel).}
\label{2d_dens}
\end{figure}

The upper panel of Fig. \ref{2d_dens} depicts the projected distribution of bright stars for the cluster 
with $N=128$K
stars. We define bright stars to be all stars with masses larger than 90\% of the mass of turn-off stars
which are still main-sequence stars
or giants at $T=12$ Gyrs. Their density distribution should be representative of the distribution of light in
the cluster.
The projected density distribution of bright stars does not show a central rise and can instead be fitted 
by a model
with a constant density core according to
\begin{equation}
 \rho=\frac{\rho_0}{(1+r/a)^5}
\end{equation}
where $\rho_0$ and $a$ are constants. A cluster with a massive central black hole would therefore
appear as a standard King profile cluster to an observer, making it virtually indistinguishable from
a star cluster before core collapse. 
Core collapse clusters have central density profiles that can be fitted by power-laws with slopes of 
$\alpha \approx 0.7$ \citep{Luggeretal1995}, which is in contradiction with this profile.
Since the central relaxation times of core collapse
clusters are much smaller than a Hubble time, any cusp profile would have been transformed into a
constant density core if an IMBH would be present in any of these clusters (see Fig.\ 5). The presence of IMBHs in
core collapse clusters is therefore ruled out unless their composition is radically different
from our clusters.

The lower panel of Fig.\ \ref{2d_dens} shows the velocity dispersions, both
the measured one and the one inferred from the mass distribution of stars.
The inferred velocity dispersions were calculated from Jeans equation (Binney \& Tremaine 1986, eq.\ 4-54)
and different mass distributions under the assumption
that the velocity distribution is isotropic (i.e. $\beta=0$). The velocities calculated from the mass
distribution of the cluster stars alone give a good fit at
radii $r/r_h>0.2$ where the mass in stars is dominating
(except at the largest radii, where the velocity distribution becomes radially anisotropic).
At radii $r/r_h<0.2$, the contribution of the black hole becomes important. At a radius $r/r_h = 0.01$, the
velocity dispersion is already twice as high as the one due to the stars alone. For a 
globular cluster at a distance of a few kpc, such a radius corresponds to central distances of one 
or two arcseconds. Of order 20 stars
would have to be observed to detect the central rise at this radius with a 95\% confidence limit. This 
seems possible both for radial velocity \citep{Gerssenetal2002} or proper motion studies
\citep{McNamaraetal2003} with HST. For a nearby globular cluster, the detection of a massive central
black hole through kinematical studies is therefore possible. Similarly, 
\citet{DrukierBailyn2003} concluded that the IMBH can be found by studying the tail of the velocity distribution 
through proper motions.

\begin{figure}[tb!]
\plotone{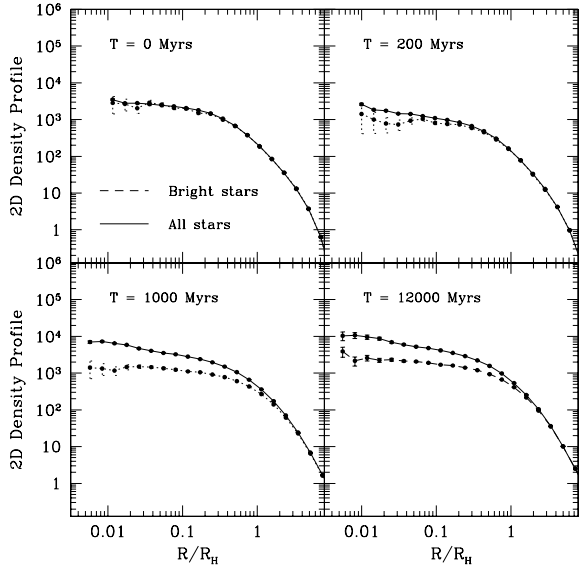}
\caption{Projected density distributions of stars for four different times for the cluster with
$N=131,072$ stars. The densities of bright stars and of all stars are shifted to match each other at
$R=10 R_H$. The density distribution of all stars develops a cusp profile after a few 100 Myrs.
Bright stars show a constant density core throughout the evolution.}
\label{rho2d}
\end{figure}

Fig.\ \ref{rho2d} shows the projected density distribution of stars for the $N=128$K cluster at
four different times. In the beginning, the
cluster has a constant density core due to the initial King model. As the cluster evolves, a central cusp 
forms around the black hole. This cusp is visible in projection already after $T=200$ Myrs and is fully 
developed around $T=1$ Gyrs. Since both times are much smaller than a Hubble time,
globular clusters which formed an IMBH early on in their evolution have reached an
equilibrium profile in their centre. Throughout the evolution, the distribution of bright stars 
displays a constant 
density core since mass segregation leads to an enhancement of heavy-mass compact remnants in the centre.
Except for the very first phases after IMBH formation when the density distribution might differ from the King
profiles with which we started, clusters with massive black holes should exhibit constant 
density cores in their light profile.

\subsection{Mass segregation}

Due to relaxation, massive stars sink into 
the centre of a star cluster in order to achieve energy equipartition between stars of different
masses \citep{Spitzer1969}. \citet{BaumgardtMakino2003} and \citet{Gurkanetal2004} showed that mass segregation of
high-mass stars proceeds on the same timescale as the evolution of the clusters toward core collapse. 
\citet{BaumgardtMakino2003} also found that by
the time a globular cluster goes into core collapse, the majority of stars in the core are compact
remnants.

\begin{figure}[tb!]
\plotone{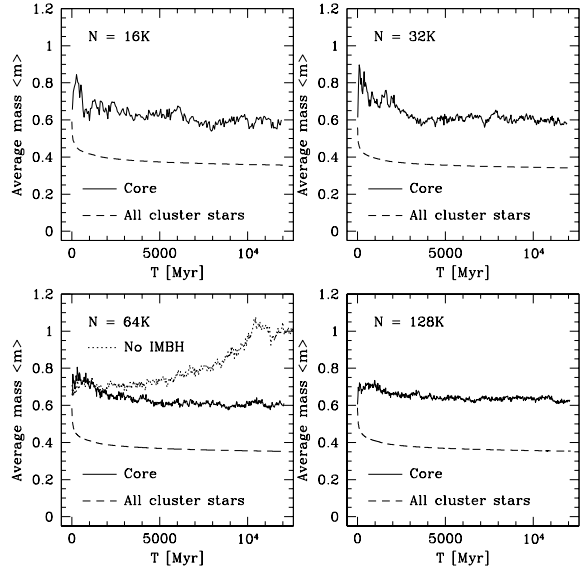}
\caption{Average mass of stars in cluster simulations with different particle numbers. In a cluster  
 without an
IMBH (dotted line, N=64K) the average mass in the core increases until it reaches a maximum at
core-collapse time (T = 10.5 Gyrs). In contrast, the average core mass for clusters with IMBHs
stays nearly constant throughout the evolution at $<\!\!m\!\!>=0.6$, independent of $N$.}
\label{mseg1}
\end{figure}

Fig.\ \ref{mseg1} shows the average mass of stars in the core (defined to contain the innermost 3\% of
the cluster mass) and the average mass of all cluster stars for the first 4 clusters of Table 1.  
The average mass of all cluster stars (dashed lines)
drops due to stellar evolution, which is most effective within the first Gyr. For low-$N$
models the core mass rises initially since the relaxation
time is short enough to allow heavy main sequence stars to spiral into the core. For higher-$N$
models, the relaxation time is larger than the stellar evolution time and stars with masses
$M>5 M_\odot$ turn into compact
remnants and lose a large fraction of their mass before reaching the core.

For N=64K stars, we
also performed a comparison run which started from a King $W_0=7.0$ model with the same IMF as the other 
clusters,
but did not contain an IMBH (dotted lines in lower left panel). 
The average mass of stars in the core for the $N=64$K cluster without a black hole rises as the cluster evolves towards
core collapse (reached at T=10.5 Gyrs). Due to the choice of IMF, there are only a few black holes
present in this model, so by the time core collapse has been reached the
average mass of stars in the core is roughly equal to the mass of the most heavy white dwarfs and 
neutron stars. This mass stays nearly constant in the post-collapse phase.
\begin{figure}[tb!]
\plotone{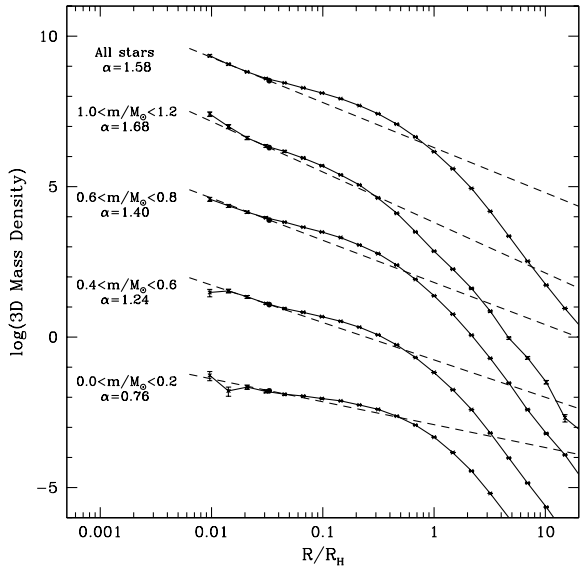}
\caption{Density of different mass groups as function of radius for the $N=128$K star cluster
at
T=12 Gyrs. Stars in the most massive group follow a distribution that drops as $\alpha \approx 1.75$
in the cusp around the black hole. For lower mass stars, the cusp profile becomes
increasingly flatter. The overall mass density drops with an average value near $\alpha=1.55$
since too few heavy mass stars are in the cusp to dominate the profile.}
\label{mseg2}
\end{figure}

In contrast, in clusters with intermediate-mass black holes the average mass of stars in the core reaches 
$<\!\!m\!\!>=0.6 M_\odot$
after 2-3 Gyrs and stays more or less constant afterwards.
There is no dependence of the average core mass on the particle number and no time evolution.
This implies that our clusters have reached an equilibrium state in which
heavy stars sink into the cusp due to mass segregation and are expelled equally rapidly from the 
cluster centre
by a balancing process.
The most likely driving force for this balancing process are close encounters between stars in the cusp.
In paper I it was shown that close encounters are efficient in removing stars from the
cusp since the average velocities are high, so stars scattered out of the cusp leave the cluster 
completely or are scattered into the halo and take a long time reaching the core again. 
The average stellar mass in the core is similar for star clusters with IMBHs and pre-collapse star 
clusters without IMBHs, making a distinction between both types of clusters through star counts difficult. 

Fig.\ \ref{mseg2} shows the density distribution of stars of different mass groups for the
cluster with $N=131,072$ stars.
Within the uncertainties, density distributions of different mass groups can be fitted 
by power-laws. The exponents $\alpha$ decrease from the heavy-mass to the lighter stars since mass 
segregation enhances the density of heavy-mass stars in the centre. 
If we exclude the few stellar-mass black holes, the most massive
stars are massive white dwarfs which
have masses $m < 1.2 M_\odot$. Fig.\ \ref{mseg2} shows that their density distribution follows a power-law 
with exponent
$\alpha=1.68$. For the $N=32$K and $64$K clusters we obtain similar exponents of $\alpha=1.80$ and 
$\alpha=1.82$ for the heavy mass stars. Given the error with
which $\alpha$ can be determined, all values are probably compatible with an exponent of $\alpha=1.75$.  
\citet{BahcallWolf1977} showed that in a two-component system, stars of 
the heavier mass group
follow a power-law distribution with $\alpha=1.75$ in the cusp around the black hole. Our simulations
are an extension of their work to multi-component systems. Here again the heaviest stars 
follow an $\alpha=1.75$ law. Since the mass in the heaviest mass group is only a small fraction of the
total cusp mass, the actual density distribution of the cusp is flatter than an $\alpha=1.75$ power-law.
The slope $\alpha$ for stars of average mass $m$ can be approximated by: 
\begin{equation}
 \alpha_{(m)} = 0.75+m/1.1
\end{equation}
We find that this density law is a good fit to all runs.
\citet{Tremaineetal1994} showed that the slope $\alpha$ of the stellar density distribution around 
a black hole
has to fulfil the condition $\alpha>1/2$ if the velocity distribution is isotropic, since otherwise 
the average velocity of stars at a given radius $r$ in the cusp would become larger than the escape velocity 
at this radius.
The density profile of the lowest mass stars is indeed above this limit even if the
argument is not a strict argument for our case since Tremaine et al. considered only the escape
from the potential of the black hole while here the main cluster mass is in the form of stars.

\subsection{Tidal disruption of stars}

\begin{figure}[tb!]
\plotone{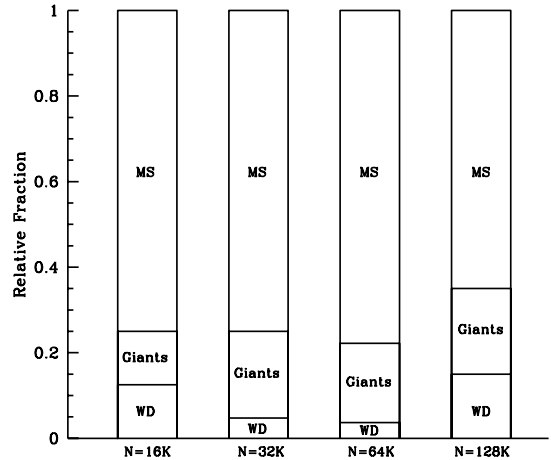}
\caption{Relative fraction of disrupted stars for cluster simulations with 
  different particle numbers. IMBHs in globular clusters preferentially
  disrupt main-sequence stars and giants, with white dwarfs accounting for only 10\% of all 
   disrupted stars.
   The number of disrupted neutron stars is negligible small.}
\label{tdis}
\end{figure}

Fig.\ \ref{tdis} shows the relative fraction
of disrupted stars
in the simulations for clusters with particle numbers between $16,384 \le N \le 131,072$.
The stars in our simulations fall roughly into four different categories: main-sequence
stars, giants, white dwarfs and neutron stars and black holes.
Mostly main-sequence stars are disrupted since they are abundant and have relative large radii.
Giants also contribute significantly despite their low numbers. Our simulations show that most giants
are disrupted within the first Gyr. There are no significant differences in the disruption rates between
individual runs. Taking the mean over all simulations,
we find that  $72\% \pm 7\%$ of all disrupted stars are main-sequence stars,
$19\% \pm 3\%$ giants and $9\% \pm 2\%$ white dwarfs. During our simulations no neutron stars or black
holes
merged with the central IMBH. However, the merging rate of neutron stars and black holes could be underestimated
since our simulations did not include gravitational radiation, the effects of which will be
discussed in section 3.6.

Using the loss-cone theory developed by \citet{FrankRees1976}, one can show with an argumentation similar
to the one in paper I that the rate at which 
stars are disrupted by a central black hole is proportional to
\begin{equation}
 D  \sim \sqrt{G} \,\left(\frac{r_T^{9-4\alpha} \, n_0^{7} \, 
   m^{4\alpha-2}}
          {M_{BH}^{5\alpha-6}}\right)^{1/(8-2\alpha)}
\label{dis1}
\end{equation}
where $r_T$ is the tidal radius of stars, $m$ their average mass and $n_0$ a constant which describes the number 
density of stars in the central cusp around the black hole $n(r) = n_0 r^{-\alpha}$. With the help of  
eq.\ \ref{koch},
we can rewrite this as:
\begin{equation}
 D  = k_D \sqrt{G} \,\left(\frac{R_*^{9-4\alpha} \, n_0^{7} \, m^{\frac{16}{3}\alpha-5}}
 {M_{BH}^{\frac{19}{3}\alpha-9}} \right)^{1/(8-2\alpha)} \;\; .
\label{dis2}
\end{equation}
In globular clusters, the radii and masses of stars differ, so disruption rates have to be calculated 
separately for each stellar
type. 

By comparing the actual number of disruptions and the estimated number
from eq. \ref{dis2}, we can determine the coefficient $k_D$. In order to do this,
we calculated $n_0$ for all times data was stored from the stars inside the cusp, and integrated eq.\ 
\ref{dis2} over time to obtain the expected number of disruptions for each run and each stellar species.
In order to calculate $n_0$, we assumed $\alpha=1.55$ for main sequence stars and giants and 
$\alpha=1.75$ for compact remnants. 
The disruption constants $k_D$ and the amount by which different stellar species should contribute
can be found in Table 2. Most simulations are compatible with $k_D=65$, which was also found in
the single-mass runs of paper I.
In agreement with the simulations, we expect most disrupted stars to be 
main-sequence stars. Giants and white dwarfs should account for about 25\% 
of all disruptions and the rate at which neutron stars are disrupted is 
negligible, in
agreement with the fact that no such disruptions were observed in our $N$-body runs.
The relative fraction of disrupted giants and white dwarfs agrees very well with the results of our runs.
Most stars are disrupted within the first two Gyrs, when the clusters are compact and the densities
around the IMBHs are high.
\begin{table}
\caption[]{Merger rates of different stellar species.}
\begin{tabular}{rrrrrr}
\noalign{\smallskip}
 \multicolumn{1}{c}{$N$}& \multicolumn{1}{c}{$k_D$} & \multicolumn{1}{c}{$f_{MS}$} &
   \multicolumn{1}{c}{$f_{Giant}$}& \multicolumn{1}{c}{$f_{WD}$} & \multicolumn{1}{c}{$f_{NS}$} \\
 16384 & $65 \pm 11$ & 0.76 & 0.15 & 0.08 & $1\cdot10^{-8}$ \\ 
 32768 & $83 \pm 18$ & 0.76 & 0.15 & 0.08 & $5\cdot10^{-9}$ \\
 65536 & $59 \pm 12$ & 0.74 & 0.17 & 0.09 & $2\cdot10^{-9}$ \\
131072 & $46 \pm 7$  & 0.73 & 0.20 & 0.07 & $2\cdot10^{-9}$ \\
\noalign{\smallskip}
 32768 & $58 \pm 41$ & 0.88 & 0.04 & 0.08 & $3\cdot10^{-5}$ \\
 65536 & $63 \pm 27$ & 0.84 & 0.04 & 0.11 & $5\cdot10^{-6}$ \\
 131072& $86 \pm 23$ & 0.83 & 0.09 & 0.08 & $3\cdot10^{-6}$ \\
\end{tabular}
\begin{flushleft}
\end{flushleft}
\end{table}

\begin{figure}[tb!]
\plotone{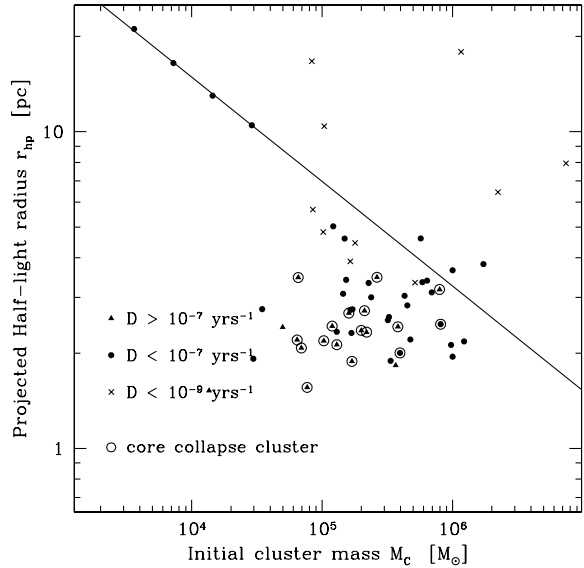}
\caption{Disruption rates $D$ for globular clusters from the list of \citet{PryorMeylan1993} for which 
 central velocity dispersions and densities are available. Core-collapse clusters have high disruption rates but
  their morphology shows that they cannot contain IMBHs. All clusters which follow the mass-radius relation
  predicted by our runs have only small disruption rates, making it unlikely that the IMBHs would be 
   visible as X-ray sources.}
\label{dcl}
\end{figure}

In order to apply our disruption rates to real globular clusters, for which the density
$n_0$ of the central cusp is unknown, we have to connect $n_0$
to the core density $n_c$. We assume that the cusp density rises with $\alpha=1.55$ and
that the cusp goes over into a constant density core with density $n_c$ at the influence
radius of the black hole, given by eq.\ \ref{ri}.
With this, the disruption rate becomes
\begin{eqnarray}
 \nonumber D  & = & \frac{1.13}{100 \mbox{Myrs}} \left(\frac{R_*}{R_\odot}\right)^{\frac{3}{5}} 
   \left(\frac{m}{M_\odot}\right)^{\frac{3}{5}}  \left(\frac{M_{BH}}{1000 M_\odot}\right)^2 \\ 
   & &  \left(\frac{n_c}{10^5 \, \mbox{pc}^{-3}}\right)^{\frac{7}{5}}  
     \left(\frac{v_c}{10 \, \mbox{km/sec}}\right)^{-\frac{21}{5}}
\label{dis3}
\end{eqnarray}
Using this formula and the velocities and central densities of globular clusters given in \citet{PryorMeylan1993}, 
we can calculate the disruption rates $D$ for globular clusters.
We assume an average stellar mass in the core of $m=0.6 M_\odot$ and a stellar radius 
$R_*=0.7 R_\odot$. The results are shown in Fig.\ \ref{dcl}. Core-collapse clusters have $D$ values of up to several
$10^{-5}$ yr$^{-1}$ due to their high central densities. However, their morphology rules out intermediate-mass black 
holes, as was shown in section 3.2. Among
the non-core collapse clusters, three clusters (NGC 1851, NGC 4147 and NGC 6336) have $D >10^{-7}$ yr$^{-1}$, so a 
1000 $M_\odot$ IMBH could double its mass within a Hubble time. The disruption rate is also high enough that X-ray 
flares from material falling onto the central black hole could be observable, depending how long it takes for the
material to be swallowed by the central black hole. These clusters are, however, more concentrated than
what would we expect if they contain intermediate-mass black holes.
Clusters which have sizes that agree with our predicted ones have fairly 
small disruption rates of $D<10^{-9}$ yr$^{-1}$, i.e. less than one star per Gyr. The black holes in such clusters
would therefore remain inactive for most of the time.
If we start from an 50 $M_\odot$
seed black hole, the disruption rates would drop to less than $2.5 \cdot 10^{-10}$ yr$^{-1}$ according to eq.\ \ref{dis3} 
even for the most favourite clusters.
Therefore, the growth of an IMBH from a black hole produced by normal stellar evolution through the tidal 
disruption of stars is impossible, unless the cluster is initially significantly more concentrated than any cluster
we see today.

\begin{figure}[tb!]
\plotone{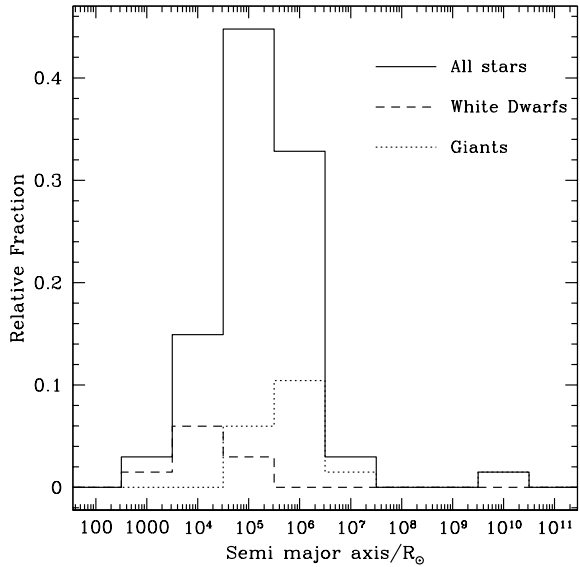}
\caption{Semi-major axis distribution $a$ of stars disrupted by the black hole in the $N=65,536$ and 
 $N=131,072$ cluster simulations.
 Most stars move on highly eccentric orbits with $a>>r_T$ when they are disrupted by the black hole. 
 Disrupted
 white dwarfs move on orbits with semi-major axes that are about a factor of 10 smaller than main-sequence stars.}
\label{merg}
\end{figure}

Fig.\ \ref{merg} shows the semi-major axis distribution of stars disrupted by the IMBH on their last orbit prior
to disruption.
Shown is the combined distribution for the $N=64$K and $N=128$K clusters which are closest to real globular clusters. All
stars have semi-major axis $a$ which are far larger than their tidal radii, similar to the situation found for
single-mass clusters in paper I. The most important mechanism for the disruption of stars is therefore again drift in 
angular momentum space. Material from disrupted stars will first move in a highly eccentric ring around the
IMBH. The ring shrinks due to viscous heating until the gas from the disrupted star
forms an accretion disc around the IMBH. Since there are other stars moving inside the initial orbit of the 
disrupted star which can scatter away material or swallow it, the fraction of material that is finally 
swallowed by the IMBH is rather uncertain.

Due to their smaller radii, disrupted
white dwarfs also come from smaller distances. According to the theory developed in paper I for single-mass 
clusters, their average semi-major axis should be smaller by about
$(R_{MS}/R_{WD})^{4/9} \approx 10$, which is in good agreement with the results in our simulations. Similarly,
giants come from larger distances on average.

\subsection{Clusters with high-mass black holes}

We now discuss the evolution of clusters which started with a mass function extending up to 100 $M_\odot$.
These clusters contained a significant number of stellar mass black holes. The highest mass black holes formed
had about 45 $M_\odot$ and we assumed a  100\% black hole retention rate, so the true fraction of black holes
in globular clusters will probably be somewhere 
between the situation in these runs and the previous ones.  Cluster radii were chosen such that
the relaxation time was the same in all
runs and equal to the relaxation time of a dense globular cluster with $M_C=10^6 M_\odot$ and $r_h=1$ pc.
\begin{figure}[tb!]
\plotone{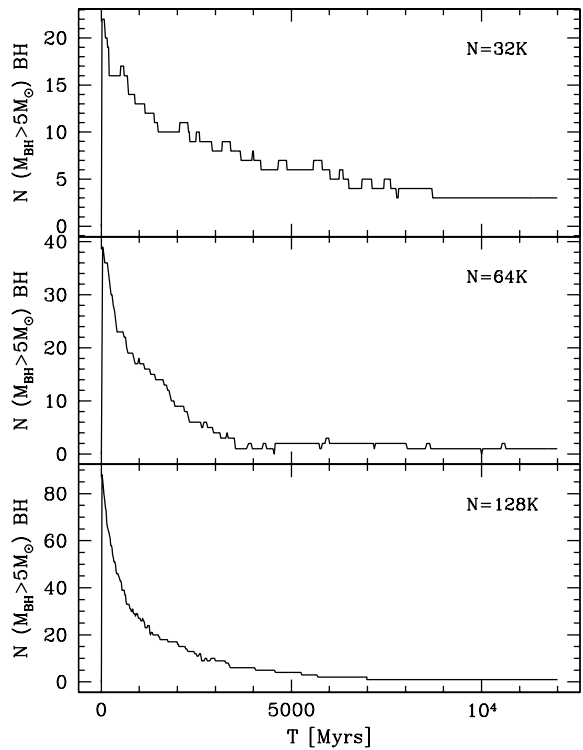}
\caption{Number of bound black holes with masses $M_{BH}>5 M_\odot$ in the runs with a high upper mass limit. The
number decreases due to close encounters between the black holes in the central cusp around the IMBH. After
several Gyrs, only one heavy mass black hole remains in most cases. This is the object most deeply bound to the
central IMBH and has absorbed the energy from the encounters which ejected the other black holes.}
\label{nhbh}
\end{figure}

Massive black holes are initially produced throughout the clusters and sink towards the centres due to mass 
segregation. Fig.\ \ref{nhbh} shows the number of black holes bound to the clusters as a function of time.
The number of heavy stellar-mass black holes drops since close encounters between
the black holes in the cusp around the IMBH remove them from the cluster. Since the velocity in the 
cusp is relatively high, such encounters
mostly lead to the ejection of one of the black holes. For the $N=64$K and $N=128$K clusters, only one massive black 
hole remains in the cluster after several Gyrs have passed. For the $N=32$K cluster 3 remained, however two of them 
were in orbits
far away from the cluster centre and would have been lost from the cluster if the cluster would have been surrounded 
by a galactic
tidal field. In all three clusters, the black hole which remains is the star closest bound to the central IMBH (see 
Fig.\ \ref{near_u}) and is among the most heavy black holes produced. This
resembles the situation in our runs with the lower upper mass limit. As the relaxation time in our runs is similar
to that of dense, massive globular clusters, we expect that clusters with IMBHs also contain only few
other massive black holes, and the star closest bound to the IMBH should be another black hole.

\begin{figure}[tb!]
\plotone{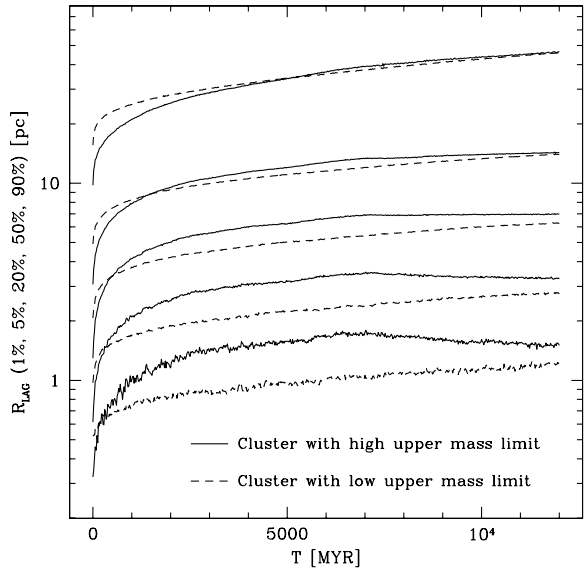}
\caption{Evolution of lagrangian radii for two $N=128$K clusters, one having an IMF extending up to $m=100 M_\odot$ and 
 many high-mass black holes and one with a lower high-mass cut-off of 30 $M_\odot$. The cluster with many black holes expands stronger
  initially due to the efficient heating of the black holes. Nevertheless, the final radii are almost the same for both clusters.} 
\label{lagcmp}
\end{figure}

A look at the lagrangian radii (Fig.\ \ref{lagcmp}) shows that the overall expansion is slightly different from the previous
case. Clusters expand more rapidly in the beginning when they still contain many heavy mass black holes.
These are very effective in scattering low-mass stars to less bound orbits while sinking into the cluster 
centre. The innermost radii of the cluster with $N=128$K stars and a high upper mass limit decrease slightly 
after $T=7000$ Myrs.
This is due to the loss of the 2nd nearest star to the IMBH through a close encounter with the innermost star (see Fig. 
\ref{near_u}). The 2nd nearest star was an efficient heat source since it was a 15 $M_\odot$ black hole 
which moved 
in a relatively wide orbit around
the IMBH, bringing it into frequent encounters with field stars. After this BH is lost, the innermost radii shrink to adjust
themselves and the energy generation rate in the centre to the size of the half-mass radius.
The final radii of clusters with heavy mass black holes are within 10\% of the radii of 
runs with the same $N$ but a lower upper mass limit (see Table 1). The initial radii were different
for $N\ge64$K stars,
but it was shown in section 3.1 that the initial radii do not influence the final radii much. We therefore conclude that
the initial mass function does not significantly affect the final radius either.

\begin{figure}[tb!]
\plotone{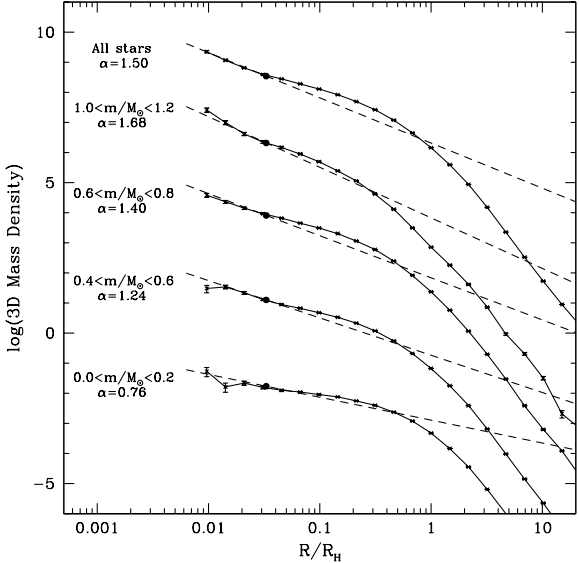}
\caption{Density of different mass groups as function of radius for the $N=128$K star cluster with a high 
 upper mass cut-off at T=12 Gyrs. The power-law slopes are similar to that of the cluster with a low upper
  mass limit. High-mass stars are again enriched in the cluster centre.}
\label{mseg_em}
\end{figure}

Fig.\ \ref{mseg_em} depicts the density distribution of the $N=128$K cluster after 12 Gyrs. As in the case
of a cluster starting with only few black holes, high-mass stars follow a steeper density distribution and
are enriched in the cluster centre. The overall slopes for stars of the same mass are very similar to the 
low upper mass limit case since most high-mass black holes have been lost from the cluster by this time, so
the overall mass function of stars is nearly identical.
Again, when viewed in projection, this cluster would appear as a cluster with a constant density core.
The average mass of stars in the core is the same as in the previous case, $<\!\!m\!\!>=0.6 M_\odot$,
independent of $N$.

The merger rates of different stars for clusters with a high upper mass limit can be found in 
Table 2. Compared to the lower mass limit case, the fraction of neutron star and black hole disruptions
is increased by a factor of 100 due to their higher density around the IMBH in the initial phases. The 
overall fraction is however still negligible. The merger rate for giants decreased 
since the high-mass stellar mass black holes prevent giants from accumulating near the IMBH in the 
initial phases. The effect is less
visible for the $N=128$K clusters, in which the disruption rate is already quite similar to the run
with the low upper mass limit. The constant for the overall disruption rate $k_D$ stayed more or less the
same, so our previous conclusions still hold if we change the mass function. This is despite the
fact that the IMBH is forced to move with a larger amplitude since the star closest bound to the IMBH is
now 20 times more massive. The movement of the central black hole has therefore not much effect on 
the merging rate.

\subsection{Gravitational radiation}

\begin{figure}[tb!]
\plotone{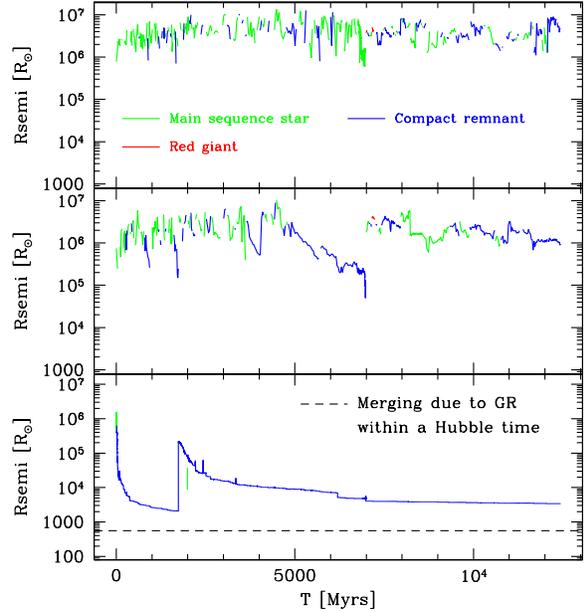}
\caption{Semi-major axis of the three stars deepest bound to the IMBH as function of time for the cluster
with $N=128$K stars and a high upper mass limit. The object closest bound to the IMBH is almost always another black hole
which is among the heaviest objects in the cluster. The other stars are too far away from the BH to undergo mass transfer.}
\label{near_u}
\end{figure}

\begin{figure}[tb!]
\plotone{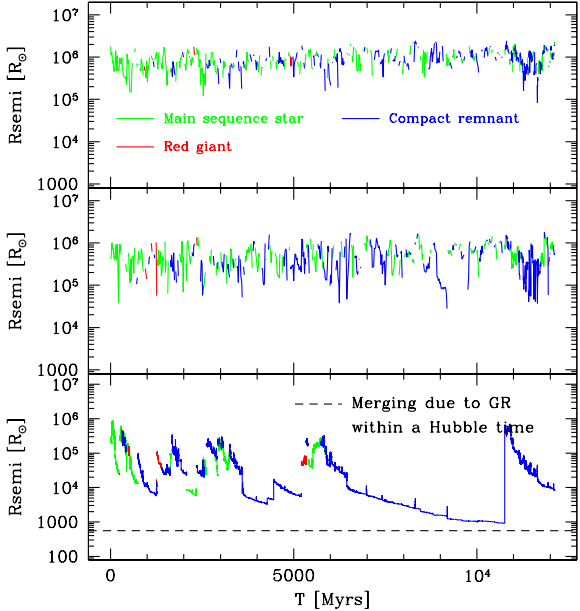}
\caption{Same as Fig.\ \ref{near_u} for the case of a low upper mass limit. In this case, the star 
 closest bound to the IMBH is
a heavy-mass white dwarf or neutron star.}
\label{near_l}
\end{figure}

Figs.\ \ref{near_u} and  \ref{near_l} depict the semi-major axis of stars which are most deeply bound to the IMBH
for the two clusters with $N=128$K stars. The energy of the deepest bound star decreases quickly in the beginning
when it still has many interactions with passing stars. When the semi-major axis becomes significantly smaller than 
that of other stars, interactions become rare and the energy change slows down considerably. In both 
clusters, the innermost
star is among the heaviest stars in the cluster, and would be a black hole with several 
$10 M_\odot$ for a globular cluster with a reasonable IMF. The innermost star will therefore not transfer mass onto the IMBH.
All other stars have 
semi-major axis of $R>10^6 R_\odot$ which is too far for mass transfer, even if some stars will move on strongly 
radial orbits. 

The time for two black holes in orbit around each other to merge due to emission of gravitational radiation 
is equal to \citep{Evansetal1987}:
\begin{eqnarray}
\nonumber T_{GR} & = & \frac{5}{256} \frac{a^4 \, c^5}{G^3 \, m_1 \, m_2 \, M} \, F^{-1}(e) \\
 \nonumber & = &\!\! 14.4 \; \mbox{yrs} \left(\frac{a}{R_\odot} \right)^4 \cdot \left(\frac{m_1}{10^3 M_\odot}\right)^{-1}
 \!\!\!\! \cdot \\ & & \left(\frac{m_2}{10 M_\odot}\right)^{-1} \!\!\!\! \cdot  \left(\frac{M}{10^3 M_\odot}\right)^{-1} 
   \!\!\!\!\!\! \cdot  F^{-1}(e)
\end{eqnarray}
where $a$ is the semi-major axis of the orbit, $c$ the speed of light, $m_1$ and $m_2$ are the masses of the two 
black holes, $M$ the
combined mass and $F$ is a function of the orbital eccentricity $e$ and is given by \citep{Peters1964}:
\begin{equation}
 F(e) = (1-e^2)^{-7/2} \left( 1 + \frac{73}{24} e^2 + \frac{37}{96} e^4 \right) \;\; .
\end{equation}
The eccentricity of the orbit of the innermost star fluctuates rapidly as long as the star undergoes many
close encounters with passing stars, but becomes fairly stable as soon as the star has detached itself from
the other cluster stars. Most of the time the innermost stars moves on an orbit with moderate eccentricity and 
we can assume $e=0.5$. With this value, the radius at which a black hole of
$20 M_\odot$ merges with a 1000 $M_\odot$ IMBH within a Hubble time is $a=562 R_\odot$. This radius is marked by 
a dashed line 
in Figs.\ \ref{near_u} and \ref{near_l}. The semi-major axis for the innermost stars in our clusters are roughly
a factor of 6 higher at the end of the simulation, so the merging timescale is around a thousand 
Hubble times, too long to have any noticeable effect on the orbits. 

\begin{figure}[tb!]
\plotone{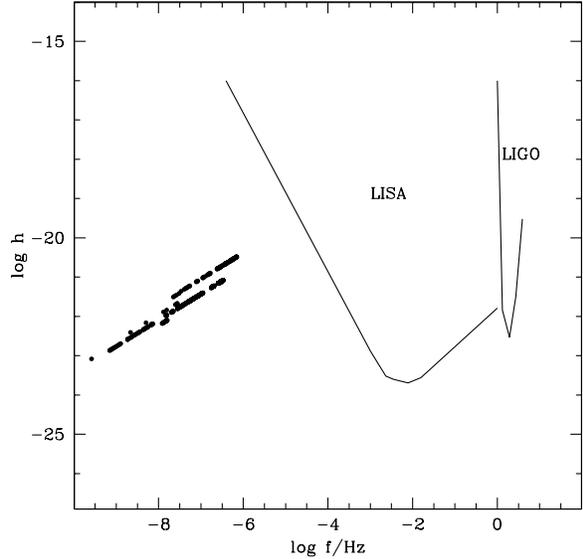}
\caption{Frequency and amplitude of gravitational radiation emitted from the innermost BH binary for the $N=128$K star
cluster which starts
with a high upper mass limit. Although the binary hardens as a result of encounters with passing stars, the
orbital separation is still too large, and hence the frequency too small, for the binary to 
become detectable.}
\label{gr}
\end{figure}

For a circular orbit, the frequency $f$ and amplitude $h$ of gravitational waves emitted by two
black holes in orbit around each other at a distance $R$ are given by \citep{DouglasBraginsky1979}: 
\begin{eqnarray}
\nonumber f & = & \frac{1}{\pi} \, \sqrt{\frac{G \, (m_1 + m_2)}{a^3}} \\
 & = &  6.2 \cdot 10^{-3} \mbox{Hz} \left(\frac{M}{10^3 M_\odot}\right)^{1/2} \!\!\!\!\! \cdot
   \left(\frac{a}{R_\odot}\right)^{-3/2}
\end{eqnarray}
and
\begin{eqnarray}
\nonumber h & = & \sqrt{\frac{32}{5}} \frac{G^2 \, m_1 \, m_2}{c^4 \, a \, R} \\
\nonumber & = &  2.63 \cdot 10^{-18} \left(\frac{m_1}{10^3 M_\odot}\right) \cdot 
  \left(\frac{m_2}{10 M_\odot}\right) \cdot  \\
  & & \left(\frac{R}{kpc}\right)^{-1} \cdot \left(\frac{a}{R_\odot}\right)^{-1}
\end{eqnarray}
Fig.\ \ref{gr} shows the amplitude and frequency of gravitational radiation emitted from the $N=128$K clusters which
started with a high upper mass limit. Each time the data were stored, we calculated both values, assuming that the cluster 
is at a distance of $R=8$ kpc. The innermost black hole is in too wide an orbit to be detectable.

How will these results change for globular clusters ? 
In clusters with higher particle numbers, the distance of the innermost stars will be different. In paper I
it was shown that the energy generation rate in the cusp is proportional to:
\begin{equation}
E_{CR} \, \sim \, G^{3/2} \, \frac{m^{3} \, n_0^{2}}{M_{BH}^{-1/2}}
\end{equation}
for an $\alpha=1.75$ cusp. Such a profile should be established very close to the IMBH in higher-$N$
models. The rate at which energy can be transfered outward at the half-mass radius is given by:
\begin{eqnarray}
\nonumber E_{TR} & \sim & \frac{N_*(r_h) \, E_*(r_h)}{T_{Rel}(r_h)}  \\
& \sim & \frac{M_C^{1.5} \, m}{r_h^{2.5}} 
\end{eqnarray}
if the weak $N$ dependence in the Coulomb logarithm can be neglected. Since for a cluster evolving slowly along a
sequence of equilibrium models energy generation must balance energy transport, a condition for $n_0$ can be
obtained from both equations. The distances of the innermost stars from the IMBH in an $\alpha=1.75$ cusp
follow a relation $r_i \sim n_0^{-4/5}$. We therefore obtain for the distance of the
innermost star from the IMBH:
\begin{equation}
r_i \sim r_h \, \frac{m^{4/5}}{M^{3/5}_C \, M^{1/5}_{BH}}
\end{equation}
Using the relation found for $r_h$ in section 3.1, $r_h \sim M_C^{-1/3}$, we obtain
for the dependence of $r_i$ on the cluster mass $M_C$:
\begin{eqnarray}
r_i \sim M_C^{-0.93}
\label{ri2}
\end{eqnarray}
i.e. the distances of the innermost stars decrease nearly linear with the particle number. Assuming an
$\alpha=1.55$ cusp gives nearly the same scaling law.
The data in our $N$-body runs is consistent with this relation. Eq.\ \ref{ri2} predicts that
even in the most massive globular clusters, the distance of the innermost main-sequence stars from the
IMBH is larger than $10^3 R_\odot$, 
too large for stable mass transfer. Dynamical evolution alone is therefore not strong enough to form stable X-ray binaries
involving the IMBH. 

As \citet{Hopmanetal2004} have shown, it might be possible
for an IMBH to capture
a passing star through tidal heating. In this case the star could end up in a circular orbit with small enough radius
so that mass transfer onto the IMBH is possible. Encounters with passing stars will however scatter the star either
out of the cusp or onto a highly eccentric orbit where it is disrupted by the IMBH. 
The present paper does not
include orbital changes due to the tidal heating of stars, so detailed simulations have to be
done to study the tidal capture rate of an IMBH and the further orbital evolution of the captured stars. 
Apart from stars, the IMBH could also accrete cluster gas lost from post-main sequence stars through stellar winds.
The resulting X-ray flux depends on the gas fraction in the cluster and the details of the accretion mechanism,
but could be bright enough to be observable under favourable conditions \citep{Hoetal2003}.

The situation looks more promising for gravitational radiation. A factor of 10 increase in the particle number from our largest
$N=128K$ star runs would already be enough that a tight IMBH-BH binary forms which
merges within a Hubble time. If the semi-major axis of a 10 $M_\odot$-1000 $M_\odot$ BH binary in a galactic 
globular cluster is less than about 50 $R_\odot$, it would become visible.
By this time, the 
time remaining to final merging has dropped to $10^8$ yrs. Making the conservative assumption that each globular 
cluster with an IMBH goes only once through such a phase, and assuming that 10\% of all galactic globular clusters
contain IMBHs, chances are around 10\% that any galactic globular cluster is presently emitting detectable 
amounts of gravitational radiation. If a GW source has a signal-to-noise level larger than 2, {\it LISA} will
have an angular resolution of a few degrees, making it possible to identify the globular cluster
containing the IMBH.

The detection is even more likely if we consider extragalactic globular clusters.
When the semi-major axis has dropped to $a=1 R_\odot$, a 10 $M_\odot$-1000 $M_\odot$ BH binary would be bright
enough to be visible to distances of $R \approx 1$ Gpc. Based on the 2dF galaxy redshift survey, 
\citet{Norbergetal2002} estimated a luminosity density of $\rho_L=(1.8\pm0.17) \cdot 10^8 L_\odot$ Mpc$^{-3}$
at $z=0$, which corresponds to roughly 0.1 Milky Way sized galaxies every Mpc$^{-3}$.
Assuming that each of them contains 100 globular clusters
gives $5 \cdot 10^{10}$ globular clusters inside 1 Gpc. Assuming again that 10\% of all globular clusters
contain an IMBH and that the merging rate of black holes with the central IMBH is constant over time, 
gives 5 events that would be visible with {\it LISA} at any given time. Globular clusters containing IMBHs
will therefore be an important source of GW emission from star clusters, in addition to double compact stars
\citep{Benacquistaetal2001}.

\section{Conclusions}

We have performed a set of large $N$-body simulations of multi-mass star clusters 
containing intermediate-mass black holes.
Our simulations include a realistic mass spectrum of cluster stars, mass-loss due to stellar evolution, two-body 
relaxation and tidal disruption of stars by the central black hole. These simulations are the first
fully self-consistent simulations of realistic star clusters with IMBHs.

Our results can be summarised as follows: A density cusp forms around the central black hole with a density profile
$\rho \sim r^{-1.55}$ in three dimensions. For low-mass IMBHs with a mass less than a few percent of the 
cluster mass, the cusp extends out to a radius where the velocity dispersion in the core becomes comparable to the 
circular velocity of stars around the black hole. In this case, the stars in the cusp contain only a fraction
of the mass of the central black hole, which makes the direct detection of the cusp difficult for IMBHs of
$M_{BH} \leq 1000 M_\odot$.  
Globular clusters with IMBHs following the relation found by \citet{Gebhardtetal2000} for galactic bulges belong
to this category. Only more massive black holes create a power-law cusp profile throughout the cluster
core that would be directly visible. 

When viewed in projection, the luminosity profiles of clusters with massive black holes 
display a constant density
core. The presence of intermediate-mass black holes in galactic core-collapse clusters like M15 is 
therefore ruled out unless these clusters have a stellar mass distribution very different from our clusters.
As was shown in \citet{Baumgardtetal2003a}, a more natural explanation 
for mass-to-light ratios that increase toward the centre in such clusters is a dense concentration of 
neutron stars, white dwarfs and stellar mass black holes.
The amount of mass segregation in a cluster with an IMBH is also smaller compared to a post-core collapse
cluster, the average mass of stars in the centre being about $m=0.6 \, M_\odot$. Clusters with IMBHs therefore
resemble star clusters which are in the pre-collapse phase also in terms of the amount of mass segregation.

All clusters with intermediate-mass black holes expand due to close encounters of stars in the cusp around the 
central black hole. We find that the values of the half-mass radii reached depend on the mass of the central black hole and 
the number of cluster stars, but are nearly independent of the initial cluster radius and density profile.
\citet{PortegiesZwartetal2004} have shown that central densities of more 
than $10^6 
M_\odot/$pc$^3$ are necessary to form an IMBH through runaway merging of massive main-sequence stars. Similarly
high densities are necessary to form an IMBH through the merging of stellar mass black holes through gravitational
radiation \citep{MouriTaniguchi2002}. Such densities are among the highest found in globular clusters. Our simulations 
show that even if clusters with intermediate-mass black holes start with very high densities, the subsequent cluster
expansion is sufficient to put them among the least concentrated clusters after a Hubble time. 
Low-mass clusters
surrounded by a strong tidal field will dissolve due to the cluster expansion, releasing their IMBHs. For clusters
close enough to a galactic centre, these IMBHs could then spiral into the centre and merge through the emission of 
gravitational radiation. If enough IMBHs are formed, this process might provide the seed black holes for the supermassive
black holes observed in galactic centres \citep{Ebisuzakietal2001}.

IMBHs in star clusters disrupt mainly main-sequence stars and giants, with white dwarfs
accounting for only 10\% of all disruptions. In young star clusters, the tidal disruption rate of giants
is similar to that of main-sequence stars. During our simulations, no neutron stars were disrupted, so intermediate-mass
black holes in star clusters do not emit gamma rays or gravitational radiation from such events. Most 
stars that were disrupted moved around the central black hole on highly
eccentric orbits with large semi-major axis. Even if 100\% of the mass
from disrupted stars is being accreted onto the central black hole, tidal disruptions of stars are too rare
to form an intermediate-mass black hole out of a $M_{BH} \sim 50 M_\odot$ progenitor,
except the initial
cluster was significantly more concentrated than present day globular clusters. In such cases, the density would however
also be high enough that run-away merging of heavy mass main-sequence stars would lead directly to the formation of an IMBH.
The largest disruption rates for globular clusters which do not have core-collapse profiles are about 
$10^{-7}/$yr for clusters with
half-mass radii significantly smaller than predicted by our runs. Clusters with
half-mass radii in agreement with our simulations have two orders of magnitudes smaller disruption rates. The
black holes in such clusters are therefore inactive for most of the time.

The detection of a $1000 M_\odot$ IMBH in a globular cluster through the measurement of radial velocities or
proper motions of cluster stars requires the observation of about 20 stars in the central cusp. For 
globular clusters that are close enough, the central cusp extends to distances of several arcsec, so the detection of an
intermediate-mass black hole should be possible by either radial velocity or proper motion studies with HST.

Black holes with masses of $5 M_\odot$ or higher are strongly depleted in star clusters with intermediate-mass black 
holes since they sink into the centre through dynamical friction and then remove each other by close encounters in
the central cusp around the IMBH \citep{Kulkarnietal1993}. Based on our simulations, we expect that only one high-mass black hole remains in 
the cluster. This black hole is among the heaviest black holes formed and ends
up as the object most tightly bound to the IMBH. In our runs, the distance of the innermost black hole to the IMBH
was never small
enough that the frequency of gravitational radiation was in the range observable for e.g. {\it LISA}.
For higher particle numbers or for clusters that start off more concentrated, it seems likely that a tight enough BH-IMBH binary 
is formed dynamically. If 10\% of all globular clusters contain IMBHs, and each IMBH merges with a stellar-mass
black hole at least once within a Hubble time, 
the chance that any galactic globular cluster currently has a tight enough BH-IMBH binary detectable for 
{\it LISA} is 10\%. Within a radius $R=1$ Gpc, about 5 globular clusters would harbour bright enough sources for {\it LISA}
at any one time.

\section*{Acknowledgements}
We are grateful to Sverre Aarseth for making the NBODY4 code available to us and his constant help with the
program. We also thank Piet Hut, Marc Freitag and Simon Portegies Zwart for useful comments. We finally thank the
referee, Fred Rasio, for a careful reading of the manuscript and comments that lead to an improvement of
the paper.

\end{document}